\documentclass{vow2008}

\title{The BMW-Chandra survey. Serendipitous Source Catalogue}

\author{P. Romano}
\affil{INAF, Istituto di Astrofisica Spaziale e Fisica Cosmica, Via U.\ La Malfa 153, I-90146 Palermo, Italy}
\author{R. P. Mignani}
\affil{Mullard Space Science Laboratory, University College London, Holmbury St. Mary, Dorking, Surrey, RH5 6NT, UK}
\author{S. Campana, A.\ Moretti, M.R.\ Panzera, G.~Tagliaferri}
\affil{INAF, Osservatorio Astronomico di Brera, Via E.\ Bianchi 46, I-23807 Merate, Italy}
\author{M.\ Mottini}
\affil{European Southern Observatory, Schwarzschild Stra\ss e 2, 85740 Garching bei M\"unchen, Germany}

\usepackage{graphicx}
\begin{document}


\maketitle

\begin{abstract}
		We present the BMW-{\it Chandra}  source catalogue 
derived from  {\it Chandra}  ACIS-I observations (exposure time $>$ 10\,ks) public as of March 2003 by using a wavelet detection algorithm (Lazzati et al.\ 1999; Campana et al.\ 1999). The catalogue contains a total of 21325 sources, 16758 of which are serendipitous.  Our sky coverage in the soft band (0.5--2\,keV, S/N =3)  is $\sim 8$ deg$^2$ for  $F_{X} \ge 10^{-13}$  erg cm$^{-2}$ s$^{-1}$,
and $\sim 2$ deg$^2$ for  $F_{X} \ge10^{-15}$ erg cm$^{-2}$ s$^{-1}$.		
The catalogue contains information on positions,  count rates (and errors) in three energy bands. 
(total, 0.5--7\,keV; soft, 0.5--2\,keV; and hard, 2--7\,keV),  and in four additional energy bands, 
SB1 (0.5--1\,keV), SB2 (1--2\,keV), HB1 (2--4\,keV), and HB2
(4--7\,keV), as well as information on the source extension, and cross-matches with the FIRST, IRAS, 2MASS, and GSC-2 catalogues. 										
\end{abstract}

\section{Introduction}

The Brera Multi-scale Wavelet (BMW, \cite{Lazzatiea99,Campanaea99}) 
algorithm, which was developed to analyse {\it ROSAT} High Resolution Imager 
(HRI) images \cite{Panzeraea03}, was recently modified to support the 
analysis of {\it Chandra} Advanced CCD Imaging Spectrometer (ACIS) images. 
At odds with other WT-based algorithms, the BMW automatically characterises 
each source through a multi-source $\chi^2$ fitting with respect to a Gaussian model 
in the wavelet space, and has thus proven to perform well in 
crowded fields and in conditions of very low background \cite{Lazzatiea99}.
Given the reliability and versatility of the BMW,  we applied 
it to a large sample of {\it Chandra} ACIS-I  images, 
to take full advantage of  the  spatial resolution 
of {\it Chandra}  [$\sim 0.5"$ point-spread function (PSF) on-axis]. 
We thus produced the Brera Multi-scale Wavelet {\it Chandra} Survey
\cite{Romano2008} and here we  present a pre-release of this catalogue,  
which is based on a subset of the whole  {\it Chandra} ACIS observations dataset,
roughly  corresponding to  the first  3 years  of  operations.  

\begin{figure}
\vspace{-1truecm}
\includegraphics[width=8.5cm,angle=0]{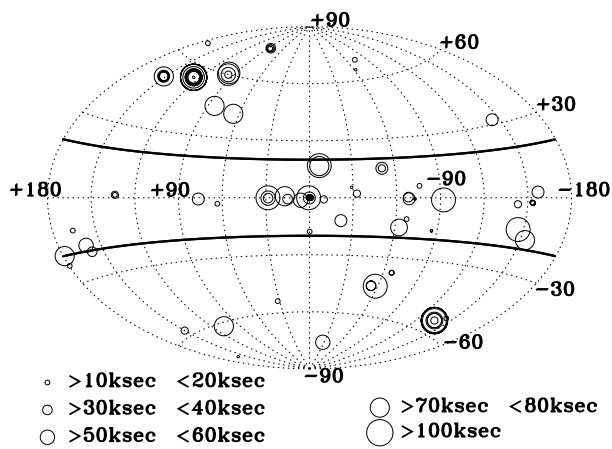}
\caption{Aitoff Projection (Galactic coordinates) of the 136
  Chandra ACIS-I fields.  The thick lines are the limits for the high latitude sub-sample.}
\label{aitoff}
\end{figure}

\section{Data selection and reduction} 

First of all, we chose the {\it Chandra} fields which maximised the sky area not centred on pointed targets. Then we selected all ACIS-I observations  with exposure time $>$ 10\,ks available by 2003 March.  Data from all four front-illuminated (FI) CCDs  were used. We excluded fields dominated by extended sources,  with bright point-like or high-surface brightness extended sources,  and fields with planets  and supernova remnants. These selection criteria  were dictated by the nature of our detection algorithm
which leads to an excessive number of spurious detections at the 
periphery of the bright source, indeed  a common problem to most detection algorithms.
Therefore, each field was visually inspected to check for such effects and, 
when found, a conservatively large portion of the field was flagged. 
Of the 147 fields analysed, 11 ($\sim 7$\%) were discarded because of problems
at various stages of the pipeline execution. We thus retained 136 fields, corresponding to 94 distinct fields, after accounting for repeated pointings. Figure~\ref{aitoff} shows the Aitoff projection in 
Galactic coordinates of their positions.  The data were retrieved from the {\it Chandra} X-ray Center (CXC) after standard processing and passed through the BMW-{\it Chandra} source detection pipeline. Briefly, the pipeline creates a WT of the input image which is then decomposed into a set of sub-images using different customised scales. For each scale, sources are identified as local maxima in the WT space  above an arbitrarily fixed significance threshold determined by the number of spurious detections computed through a Monte Carlo simulation. 
Sources from different scale images are then  cross-matched and finally characterised  through a multi-source 
$\chi ^2$ minimization with respect to a Gaussian model source in the WT space (see \cite{Romano2008} 
for a detailed description). 

\section{Catalogue description}

The WT detection pipeline produced a catalogue of source positions, count rates,  counts, extensions, and relative errors in three energy bands (total, FB, 0.5--7\,keV; soft, SB, 0.5--2\,keV; and hard, HB, 2--7\,keV). To these, we added additional information derived from the headers of the original files. 
We  extracted source counts within a box centered around the positions determined with the detection algorithm, with a side which is the 90\% encircled energy diameter at 1.50\,keV. We also extracted source counts in four additional bands: 
SB1 (0.5--1.0\,keV), 
SB2 (1.0--2.0\,keV), 
HB1 (2.0--4.0\,keV), and 
HB2 (4.0--7.0\,keV). For the SB, HB, and FB bands the background counts were  extracted from the same box from the background image.
	We converted  the count rates in fluxes assuming a Crab spectrum, i.e.\ a power law with photon index 2.0, modified with the absorption by Galactic $N_{\rm H}$ relative to each field, and 
with a simple Crab spectrum ($N_{\rm H}=0$).  \\
To characterise the source extension, 
which is one of the main features of the WT method, one cannot simply compare 
the WT width with the instrumental PSF at a given off-axis angle. 
Thus, we use a $\sigma$-clipping algorithm which divides the 
distribution of source extensions as a function of off-axis angle 
in bins of 1' width. 
The mean and standard deviation are calculated within each bin and all sources 
which width exceeds 3$\sigma$ the mean value are discarded. 
The procedure is repeated until convergence is 
reached. The advantage of this method is that it effectively eliminates truly 
extended sources, while providing a value for the mean and standard deviation 
in each bin \cite{Lazzatiea99}. The mean value plus the 3$\sigma$
dispersion provides the line discriminating the source extension, but we conservatively classify as 
extended only sources that lie 2$\sigma$ above this limit. 
Combining this threshold with the 3$\sigma$ on the intrinsic dispersion, 
we obtain a $\sim 4.5 \sigma$ confidence level for the extension classification. 
The distribution of the source off-axis angle presents a steep increase with collecting area, and a gentler decrease with  decreasing sensitivity with off-axis angles. Differently from what found with the 
BMW-HRI catalogue \cite{Panzeraea03}, our distribution does not present 
a peak at zero off-axis due to pointed sources. 
We have also cross-correlated the BMW-{\it Chandra}  catalogue with optical, infrared, and radio catalogues like the GSC-2, 2MASS, IRAS, and FIRST to select potential candidate counterparts for X-ray source identifications. 						The full catalogue contains 21325 sources, 16834 of which are 
not associated with bright and/or extended sources, including the
pointed ones. Of these, 11124 are detections in the total band,
12631 in the soft, 9775 in the hard band; 4203 sources were only detected in the hard
band. 
The current version of the BMW-{\it Chandra} source catalogue,
(as well as additional information and data) is available at the
Brera Observatory\footnote{http://www.brera.inaf.it/BMC/bmc\_home.html}   
and at the  INAF-IASF Palermo mirror sites\footnote{http://www.ifc.inaf.it/$\sim$romano/BMC/bmc\_home.html } and through the Centre de Donn\'ees
astronomiques de Strasbourg (Vizier) and the HEASARC sites. 

\section{Catalogue Exploitation}

 It is particularly important for cosmological studies to have a sample which is 
not biased toward bright objects. To this end, we constructed the 
BMW-{\it Chandra} Serendipitous Source Catalogue that contains 16758 
sources not associated with pointed objects, by 
excluding sources within a radius of 30 arcsec from the target position. 
Among the avenues of scientific exploitation of the BMW-{\it Chandra} Serendipitous Source Catalogue are: 
i) Characterization of the sources based on X-ray colours alone; 
ii) X-ray source identification based on the multi-wavelength cross-correlations, iii) selection of sub-samples of promising sources for optical follow-up like, e.g.\ blank fields (sources without counterparts at other wavelengths) or  heavily absorbed sources (the 4203 only detected in the hard X-ray
band), 
iii) analysis of a sample of ~300 extended sources, which constitutes
a list of X-ray selected galaxy cluster candidates, to be confirmed 
optically, and 
iv) temporal and spectral variability through autocorrelation of the catalogue allows to search for long-term variability of sources observed more than once and search for periodicities in the light curves.


\end{document}